# Emittance Growth Mechanisms in the Tevatron Beams[1]

**Vladimir Shiltsev**[a*] **and Alvin Tollestrup**[a]

[a] *Fermi National Accelerator Laboratory,*
*PO Box 500, Batavia, IL, 60510, USA*
*E-mail*: shiltsev@fnal.gov

ABSTRACT: In this article we present results of emittance growth measurements in the Tevatron beams. Several mechanisms leading to transverse and longitudinal diffusions are analyzed and their contributions estimated.

KEYWORDS: Instrumentation for particle accelerators and storage rings - high energy (linear accelerators, synchrotrons)

---

[1] Work supported by Fermi Research Alliance, LLC under contract No. DE-AC02-07CH11359 with the U.S. Department of Energy

* Corresponding author

# Contents

## 1. Introduction

The Tevatron proton-antiproton collider [1] was the world's highest energy collider for almost 25 years since it began operation in December 1985 until it has been overtaken by the LHC in 2009. The aim of the Tevatron collider is to explore the elementary particle physics phenomena with centre of mass collision energies of up to 1.96 TeV. The number of events per second generated in the Tevatron collisions is given by:

$$dN_{event}/dt = L \cdot \sigma_{event} , \quad (1)$$

where $\sigma_{event}$ is the cross section for the event under study, and $L$ is the machine luminosity. The machine luminosity depends only on the beam parameters and for a Gaussian beam distribution is:

$$L = \gamma f_0 \frac{N_B N_a N_p}{4\pi \beta^* \varepsilon} H(\sigma_s/\beta^*) , \quad (2)$$

where $N_{p,a}$ is the number of particles (protons or antiprotons) per bunch, $N_B$ the number of bunches per beam, $\varepsilon$ is the average rms normalized emittances of two round beams $(\varepsilon_a+\varepsilon_p)/2$, $H(x)$ is the geometric luminosity reduction factor ("hourglass factor") which depends on the ratio of the rms bunch length $\sigma_s$ and beta-function at the collision point $\beta^*$, $\gamma$ is the relativistic factor, and $f_0$ is the revolution frequency.

Evolution of the Tevatron luminosity is discussed in detail in [2] and can be well approximated by a simple empirical fit [3]:

$$L(t) = \frac{L_0}{1+t/\tau_L} , \quad (3)$$

where the only two parameters are initial luminosity $L_0$ and initial luminosity lifetime $\tau_L$.

The luminosity integral $I=\int L dt$ – the sole critical parameter for collider experiments – depends on the product of peak luminosity and the luminosity lifetime, e.g. for a single store with initial luminosity $L_0$ and duration $T$, the integral is $I \approx L_0 \tau_L \ln(1+ T/\tau_L)$.

From Eq.(4), one gets:

$$\tau_L^{-1} = \frac{dL(t)}{L(t)dt} = |\tau_\varepsilon^{-1}| + \tau_{Na}^{-1} + \tau_{Np}^{-1} + \tau_H^{-1} \quad , \quad (4)$$

i.e., the luminosity lifetime has four constituents: the growth rates of beam emittances, the beam intensity decay rates and the hourglass factor decay rate. For the end of the Collider Run II operation with range of initial luminosities between $3.0 \times 10^{32}$ cm$^{-2}$ s$^{-1}$ to $4.0 \times 10^{32}$, the largest contribution to luminosity lifetime of about $\tau_L$=5.2-5.7 hours came from the beam emittance growth with a typical time of $\tau_\varepsilon$ ~10-13 hours. The growth is dominated by intrabeam scattering (IBS) in the proton and antiprotons bunches, with small contributions from the beam gas scattering and external noises. Beam-beam effects can lead to faster emittance blowups and significant losses but they are routinely corrected or compensated. The hourglass factor decays with $\tau_H$ ~ 70-100 hours due to IBS and smaller contribution due to the RF system noises.

## 2. Experimental Study of the Emittance Growth in 980 GeV Proton Bunches

Several beam experiments have been conducted to separate contributions of different phenomena to the emittance growth. In one of them, 15 proton bunches with various intensities were accelerated to 980 GeV – see Fig.1. The bunches also had very different emittances $\varepsilon$ varied from 2.3 to 3.6 $\pi$ mm mrad and rms bunch lengths $\sigma_z$ in the range from 1.71 ns to 2.10 ns. The bunches were left in the machine for 3.1 hours and their emittances, bunchlengths and intensities were regularly measured by the Flying Wires system and the Sampled Bunch Display (SBD) systems, correspondingly [4].

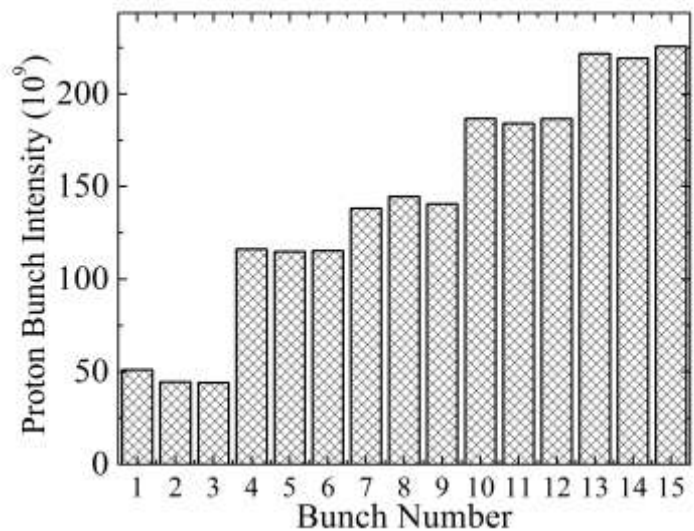

Figure 1: Proton bunch populations at the beginning of the experimental studies store.

Theory and simulations of the IBS in the Tevatron bunches where the transverse velocity spread is much larger than the longitudinal one, predicts that the time behavior of the transverse and longitudinal emittances is given by [5]:

$$\frac{d\varepsilon_T}{dt} = \frac{\gamma^{3/4} N_p C_T}{\varepsilon_T^{1.5} \varepsilon_L^{0.5}} \quad \text{and} \quad \frac{d\sigma_z^2}{dt} = \frac{N_p C_L}{\gamma^{1/4} \varepsilon_T^{1.5} \varepsilon_L^{0.5}} \quad (5),$$

$C_L$ and $C_T$ are constants determined by machine parameters, and both vertical and horizontal emittances are about the same $\varepsilon_x \approx \varepsilon_y \approx \varepsilon_T$. From (5), one expects that the IBS-driven emittance growth has to be proportional to the factor $F_{IBS} = \frac{N_p}{\varepsilon_T^{1.5} \sigma_z}$. (In our analysis we used vertical emittance as its systematic error and uncertainty for the Tevatron Flying Wires system are much smaller than for the horizontal emittance, while it is known from the experience that usually the two emittances are proportional to each other). Since the bunches were evolving in time, $F_{IBS}$ is a function of time, so we used an average value of the factor during the store and plotted the observed emittance growth and mean squared bunch length growth rates versus $F_{IBS}$ for all the bunches - see Figs.2a and 2b.

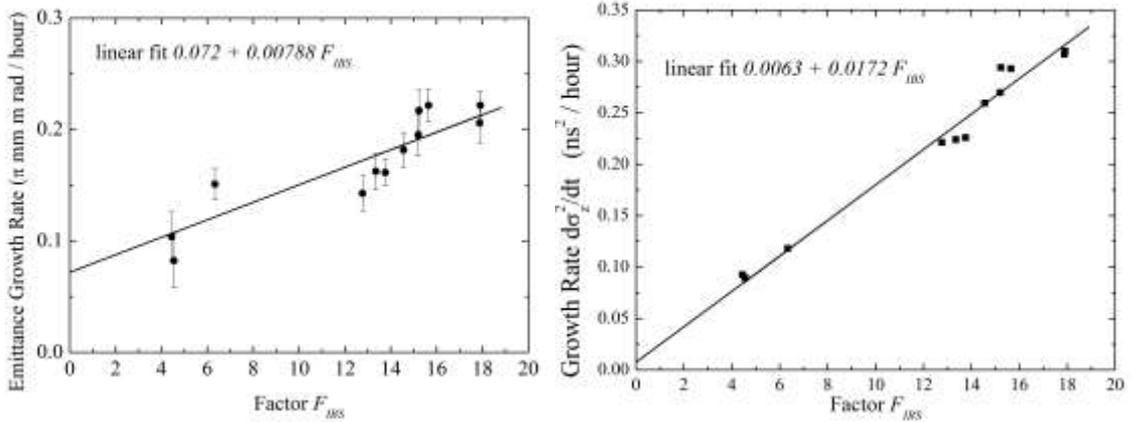

Figure 2: a) (left) Vertical emittance growth rates of proton bunches vs the IBS factor $F_{IBS}$; b) (right) the mean squared bunch length growth rates vs the IBS factor $F_{IBS}$ (see in the text).

The observed increases can be approximated by linear fits:

$$\frac{d\varepsilon_V}{dt}[\pi\, mmmrad/hr] = (0.072 \pm 0.02) + (0.0079 \pm 0.0015) \cdot F_{IBS} \quad \text{and}$$

$$\frac{d\sigma_z^2}{dt}[ns^2/hr] = (0.0063 \pm 0.0193) + (0.0173 \pm 0.0013) \cdot F_{IBS} \quad (6).$$

The intercept in the transverse emittance growth of 0.072 π mm mrad/hr gives an upper estimate on the growth due to intensity independent effects of scattering on the residual vacuum molecules and due to external noises. The emittance growth due to the gas scattering is equal to

$$\frac{d\varepsilon_{x,y}}{dt} = \frac{2\pi c r_p^2}{\gamma \beta^2}\left(\sum_i n_i Z_i(Z_i+1)L_{C_i}\right)\overline{\beta_{x,y}} \quad \text{where} \quad \overline{\beta_{x,y}} = \frac{1}{C}\int \beta_{x,y}\,ds \approx 70 \text{ m} \quad (7),$$

so, the observed zero-intensity emittance growth rate corresponds to an equivalent average room-temperature $N_2$ ($Z_i$=7) vacuum pressure of $(2.4\pm0.7)\times10^{-9}$ Torr. That is an obvious overestimate, because such a pressure would result in the beam intensity lifetime due to nuclear beam-gas interaction of about $140\pm40$ hours while in this experiment the beam lifetime was about $\tau_H \sim 800\pm500$ hours (confirmed in many other similar measurements). Therefore, one can estimate that only 20% or less of the observed low intensity beam emittance growth is due to the gas scattering – thus, corresponding to equivalent average pressure of room-temperature $N_2$ of the Tevatron is about or less than $P \approx 0.5\times10^{-9}$ Torr – and that most of the growth comes from the external noises, namely, some 0.06 π mm mrad/hr. To be noted, that large angle Coulomb gas scattering and Touchek scattering give very small contributions to beam losses, corresponding to lifetimes of about 30,000 hours and 7,000 hours, respectively.

As for the longitudinal diffusiion, from the zero intensity value in the second equation in Eq.(6) and formulae for the longitudinal emittance growth due to the RF phase noise:

$$\left.\frac{d\langle\sigma_z^2\rangle}{dt}\right|_{RF} = \pi\Omega_s^2 P_\phi(\Omega_s)\times\left(\frac{4\pi^2 c^2}{f_{RF}^2}\right) \quad (8),$$

one can estimate the power spectral density of the RF phase noise at the synchrotron frequency $f_s=\Omega_s/2\pi\approx35$ Hz to be $P_\phi(f_s) = 4\pi P_\phi(\Omega_s) \approx (1.8\pm5.5)\cdot10^{-11}$ rad$^2$/Hz. The main source of the noise is believed to be the microphonics effect in the RF cavities excited by the flow of cooling water. The RF phase feedback suppresses this noise by ~30 dB to quite acceptable level [6].

3. Discussion

Major conclusions of the 980 GeV beam studies presented above re in a good agreement with a different type of experiment conducted at the Tevatron injection energy of 150 GeV. In that experiment [7], the evolution of the transverse distribution of the low intensity un-bunched beam of protons is monitored after partial scraping of the beam, leaving a zone of high diffusion when the beam is left by itself interacting with the residual gas and blown up by external noises. The halo population immediately outside the beam core is dominated by the diffusion. The particle distribution can be deducted from the shape of the signal from the DCCT current monitor [4] when the current vanishes during vertical scraping as shown in Fig. 3a. The $x$-axis corresponds to the vertical collimator position that is pushed toward the beam from the outside of the beam pipe up to the extinction of the current (around ~28 mm). In this plot the collimator goes from right to left, and only starts to affect the beam intensity when the position has passed ~26 mm. Each point is separated by 1s, since the data acquisition was done at 1 Hz. The experiment consisted of three measurements:

1. The beam was injected and partially scraped horizontally, then totally scraped with vertical collimator, constituting the 1st reference.
2. Another injection followed with the same conditions, but the vertical scraping stopped before the removal of the beam around half the total intensity, then collimator was retracted and the beam stayed in the machine for 1.5 hours, before being fully scraped.
3. A third injection and scraping constitute the 2nd reference.

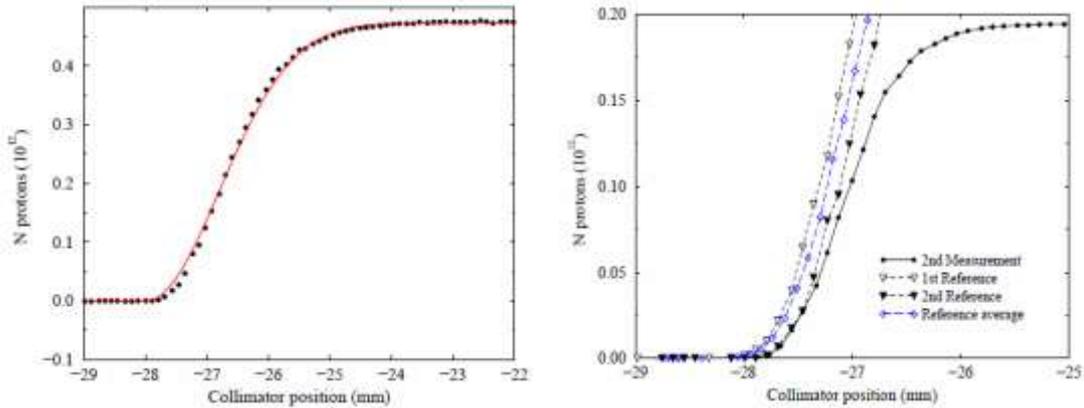

Figure 3: a) (left) beam intensity vs collimator position scan and fit curve corresponding to a Gaussian distribution; b) (right) comparison between reference and delayed scraping profiles (see text).

The emittance growth is estimated by comparing the scraping patterns for the 1st and the 2nd reference measurements with the profile after 1.5 hours (measurement 2) as shown in Fig. 3b. The evolution of the beam transverse phase-space distribution function $f$ has been modeled by the Fokker-Planck equation expressed with the action variable $I$:

$$\frac{\partial f}{\partial t} = D \frac{\partial}{\partial I}\left[ I \frac{\partial f}{\partial I} \right] \quad (9),$$

and the fit to experimental data allowed to determine the diffusion coefficient $D=(1.66\pm0.05)\times10^{-3}$ mm mrad/s, corresponding to the emittance growth rate of about $0.48\pm0.015$ $\pi$ mm mrad/hr. Given that the emittance growth rate due to both gas scattering and external noises scales as $1/\gamma$, the rate measured at 150 GeV corresponds to $(0.48\pm0.015)\times(150/980)=0.073\pm0.002$ $\pi$ mm mrad/hr at the 980 GeV – in an excellent agreement with the measurements reported above. The observed beam intensity lifetime of around $1000\pm40$ hours points to average equivalent room-temperature $N_2$ pressure of $P=0.34\times10^{-9}$ Torr (again, in a good agreement with the 980 GeV results), and, therefore, pointing out that almost three quarters of the observed emittance growth rate is due to the external noises.

## 4. Summary

Our beam experiments in the Tevatron indicate that almost all longitudinal emittance growth and at least 2/3 of the transverse emittnace growth in high intensity proton bunches are due to Intra-beam scattering effects. The remaining ~1/3 of the transverse emittance growth of about 0.07 π mm mrad/hr is mostly (~80%) due to external noises and the rest from other intensity independent effects, like scattering on residual vacuum molecules.

**Acknowledgments**

Authors are very thankful to the Fermilab's Tevatron Department staff who helped us to set up and carry out these experimental studies, to Dr.V.Lebedev for many stimulating discussions, and to Dr. L.Nicolas for taking part in the data analysis.